\documentclass[conference,a4]{IEEEtran}
\usepackage{graphicx,times,latexsym,amsfonts,amssymb,amsmath}
\usepackage{comment,url,color}

\usepackage{xspace,subfigure,epsfig}
\usepackage{graphicx,times,latexsym,amsfonts,amssymb,amsmath}
\usepackage{comment,url,color}

\newcommand{\Uno}{ \mathbb{I} }

\newcommand{\Prob} { {\rm Pr}}

\newtheorem{teorema}{\bf Theorem}
\newtheorem{corollario}{\bf Corollary}

\newcommand{\ls}[1]
{\dimen0=\fontdimen6\the\font
  \lineskip=#1\dimen0
  \advance\lineskip.5\fontdimen5\the\font
  \advance\lineskip-\dimen0
  \lineskiplimit=.9\lineskip
  \baselineskip=\lineskip
  \advance\baselineskip\dimen0
  \normallineskip\lineskip
  \normallineskiplimit\lineskiplimit
  \normalbaselineskip\baselineskip
  \ignorespaces
}

\newcommand{\tgifeps}[3]{
  \begin{figure}[t]
    \centering
    \includegraphics[width=#1.0cm]{figures/#2.eps}
    \caption{#3\label{fig:#2}}
    \vspace{-2mm}
  \end{figure}
}

\excludecomment{conference}
\includecomment{techrep}

\newcommand{\be}{\begin{equation}}
\newcommand{\ee}{\end{equation}}
\newcommand{\ba}{\begin{array}}
\newcommand{\ea}{\end{array}}

\newcommand{\youtube}{\textsc{YouTube}\xspace}

\newcommand{\RMMOA}[2]{}

\begin{document}

\begin{sloppypar}

  \title{ Analyzing the Performance of LRU Caches under Non-Stationary
    Traffic Patterns}

  \author{
    \IEEEauthorblockN{
      Mohamed Ahmed\IEEEauthorrefmark{1},
      Stefano Traverso \IEEEauthorrefmark{2},
      Paolo Giaccone\IEEEauthorrefmark{2},
      Emilio Leonardi\IEEEauthorrefmark{2} and
      Saverio Niccolini\IEEEauthorrefmark{1}
    }
    \IEEEauthorblockA{
      \IEEEauthorrefmark{1}
      NEC Laboratories Europe, Heidelberg, Germany -- \{firstname.lastname\}@neclab.eu
    }
    \IEEEauthorblockA{
      \IEEEauthorrefmark{2}
      Department of Electronics and Telecommunications,
      Politecnico di Torino, Torino, Italy -- \{lastname\}@tlc.polito.it
    }
  }

\maketitle

\begin{abstract}
  This work presents, to the best of our knowledge of the literature,
  the first analytic model to address the performance of an LRU (Least
  Recently Used) implementing cache under non-stationary traffic
  conditions, i.e., when the popularity of content evolves with time.
  We validate the accuracy of the model using Monte Carlo
  simulations. We show that the model is capable of accurately
  estimating the cache hit probability, when the popularity of content
  is non-stationary.

  We find that there exists a dependency between the performance of an
  LRU implementing cache and i) the lifetime of content in a system,
  ii) the volume of requests associated with it, iii) the distribution
  of content request volumes and iv) the shape of the popularity
  profile over time.




\end{abstract}

\section{Introduction}~\label{sec:introduction} Content caching is
today a primitive network management operation, while the operation
and performance of Content Delivery Networks (CDNs) is predicated on
understanding on how users consume content.

This argument is given urgency by two main factors. First, the
continued growth of Internet traffic, especially in the mobile
context, increases the demand on limited network
resources~\cite{cisco:2012,Sandvine:2012}. Second, the dominance of
multimedia traffic is today de-facto~\cite{Liu:2012}. Recent
studies~\cite{Sandvine:2012} show that in the US, services offering
real-time video and audio streaming occupy $62.5\%$ and $54.7\%$ of
peak-period downstream traffic for fixed and mobile networks
respectively. Similarly in Europe, real-time multimedia traffic
accounts for 33.5-50\% of peak-period downstream traffic in fixed
networks, while globally, video traffic alone is project to account
for $55\%$ of all ``consumer Internet traffic'' by
2016~\cite{cisco:2012}.

This traffic profile poses unique challenges to providing users with a
reliable Quality of Service (QoS). For instance, Liu
et. al.~\cite{Liu:2012} report that $20\%$ of users experience
re-buffering when streaming contents, while $14\%$ of users suffer
significant delays before videos start to play. Furthermore, new
networking paradigms such as ICN (Information-Centric Networking) are
built on the implicit assumption of ubiquitous content
caching~\cite{Chai:12,Rossi:12}, such that small caches are co-located
with routers in order to offset traffic latency.

Therefore, improving the effectiveness of content caching is paramount
in aiding to address the problem of scaling the network while
providing the necessary QoS to users. However, the vast majority of
the studies on caching assume traffic patterns to be time-invariant,
i.e., users browse through a large static catalogue of contents and
make requests according to their different \textit{static} popularity
distributions - typically assumed to be Zipf.

In reality however, the popularity of contents varies over time and
different contents exhibit a wide range of popularity evolution
patterns~\cite{Yang:2011, Ahmed:2013}. Contents tend to differentiate
in i) when they start to attract user attention, ii) how much
attention they attract and iii) how long they sustain the
attraction. For instance, contents related sporting or geo-political
events such as Olympic videos tend to enjoy a very short
lifetime~\cite{Sandvine:2012}, reflecting users' immediate interest in
the topic. In contrast, some \youtube music videos keep on attracting
user attention many months after being
released~\cite{Ahmed:2013}. Clearly, this observed behaviour is
incompatible with time-invariant popularity models, and raises the
need for more accurate tools that take into consideration the
evolution of the popularity of contents over time.

This work presents the first steps in this direction. We extend the
results of Che et al.~\cite{che02} to take the time-variant popularity
of contents explicitly into account, and present an approximated model
of an LRU (Least Recently Used) caching under non-stationary traffic
conditions. The accuracy of our model is validated against Monte Carlo
simulations and shows that the non-stationarity of content popularity
has a dramatic effect on caching performance. We find that even when
cache sizes are small as in the context of ICNs, i.e., when cache
dynamics change much faster than the popularity evolution, the
performance of the cache is largely sensitive to popularity dynamics.

\section{LRU under non-stationary conditions}\label{sec:modelling}
In this section, we first present the assumptions that unpin our model
(Sec.~\ref{sec:traffic}). This is followed by the derivation of the
cache hit probability under the non-stationary traffic scenario
(Sec.~\ref{sec:hit}), with specific reference to scenarios with large
(Sec.~\ref{sec:hit:large}) and small cache sizes (Sec.~\ref{sec:sc}).

\subsection{ A simple non-stationary traffic model}\label{sec:traffic}
We start by assuming that contents are introduced into a catalogue
(i.e. uploaded on some server) at random.  For simplicity, this is
taken to be according to a homogeneous Poisson process with rate
$\gamma$.  Furthermore, it is assumed that individual content
popularity evolves (over time) according to some predetermined
profile.  Initially, it is assumed that all contents follow the same
popularity profile, in Sec.~\ref{sec:ext} we show that this
assumption can be relaxed.

Let us now consider a generic content $m$, introduced into the
catalogue at time $\tau_m$, and whose popularity evolves over time
according to:
 \[
 \lambda_m(t)= V_m \lambda(t-\tau_m)
 \]
where $\lambda(t)$ represents the popularity profile and $V_m$ a
random mark (i.e. a random quantity) associated to content $m$.
Popularity ($\lambda_m(t)$) in this context represents the
instantaneous rate at which requests for a given content $m$ arrive at
the cache.

Requests are assumed to form an independent time-inhomogeneous Poisson
processes and $\lambda(t)$ is taken to be an arbitrary function
satisfying the following conditions: i) (positiveness) $\lambda(t) \ge
0$ $\forall t$ with $\lambda(0^+)>0$, ii) (causality) $\lambda(t)=0$ $\forall t<0$, iii)
(smoothness) $\lambda(t)$ continuous almost everywhere, iv)
(integrability) $\int_0^\infty \lambda(t) d t =1$.
The average content lifetime can be computed as $L=\int_0^\infty t \lambda(t)dt$.

Observe that $V_m$ represents the expected total number of requests
(volume) induced by content $m$ during its whole life in the system.
More specifically, by construction, the total number of requests for
content $m$ is given by a Poisson distribution with an average of
$V_m$.  We assume that volumes of requests for different contents form
an i.i.d.\ sequence of random variables distributed around some
reference $V$.  We denote $\phi_V(x)=\mathbb E[e^{xV}]$ to be the
moment generating function of $V$ and $\phi'_V(x)$ its first
derivative.
 
Finally, the aggregate process of requests arriving to the cache is
now by construction a Cox process~\cite{cox} whose stochastic
intensity is given by:
\(
\Lambda(t)=\sum_{m} V_m \lambda(t-\tau_m) 
\).

\subsection{Cache hit probability}\label{sec:hit}
In this section we extend the LRU approximation of Che et.
al.~\cite{che02} to our non-stationary traffic model in order to
estimate the cache hit probability, i.e.\ the probability that a
generic request finds the content in the cache.

Consider a cache capable of storing $C$ distinct contents. Let
$T_C(m)$ be the time needed for $C$ distinct contents not including
$m$ to be requested by users. $T_C(m)$ therefore represents the {\em
  cache eviction time} for content $m$, i.e. after which point content
$m$ will be evicted from the cache. Che's approximation is centred on
assuming that the cache eviction time ($T_C(m)$) is deterministic and
independent from the selected content ($m$). This assumption has been
given a theoretical justification in~\cite{Roberts12}, where it is
shown that, under a Zipf-like static popularity distribution, the
coefficient of variation of the random variable representing $T_C(m)$
tends to vanish as the cache size grows. Furthermore, the dependence
of the eviction time on $m$ becomes negligible when the content
catalogue is sufficiently large.  The arguments given
in~\cite{Roberts12} are easily extended to our non-stationary traffic
model when $\gamma$ and $C$ are sufficiently large.

Returning to our non-stationary traffic model, we can now state our
main result:
\begin{teorema}\label{theo:1}
  Consider a cache of size $C$ implementing LRU policy, operating
  under a non-stationary popularity model (as introduced in
  Sec.~\ref{sec:traffic}) with total stochastic intensity: \(
  \Lambda(t)=\sum_{m} V_m \lambda(t-\tau_m) \). Extending Che's
  approximation, the hit probability is given by:
\begin{equation}\label{eq:phit}
  p_{\text{hit}} = 1 - \int_0^{\infty}\lambda( \tau) \
  \frac{\phi_V' \left(-\int_ {0}^{T_C} \
      \lambda(\tau - \theta)d \theta \right)}  { \mathbb{E}[V] }  d \tau  
\end{equation}
where $T_C$ is the solution to the equation:
\begin{equation}\label{eq:c2}
  C=\gamma \int_{0}^\infty 1 - \phi_V \ 
  \left(-\int_ {0}^{T_C} \lambda( \tau - \theta) d\theta \right) d \tau 
\end{equation}
and $\gamma$ is the rate at which new contents are introduced into the
catalogue.
\end{teorema}
\begin{IEEEproof}
\begin{techrep}
  Proceeding along the same lines as for the stationary case 
  (recalled in Appendix~\ref{sec:static}), we consider a constant
  cache
\end{techrep}
\begin{conference}
  Proceeding along the same lines as for the stationary case~\cite{che02}, we consider a constant
  cache
\end{conference}
eviction time $T_C$.  We can now start evaluating the probability of
finding a given content in the cache at time $t$, conditional on the
time it has been introduced into the catalogue ($\tau$) and its
request volume
($V$).  This corresponds to the event in which one or more requests
for the content occur in the time interval $[t-T_C,t]$.  This
probability can be evaluated
\begin{techrep}
  as~\footnote{similarly to~\eqref{eq:phm} obtained under stationary
    popularity}:
\end{techrep}
as:
\begin{equation}\label{eq:ph2}
  p_{in}(t\mid \tau,V) = 1- e^{-V \int_ {t-T_C}^{t}\lambda(\theta-\tau)d\theta}
\end{equation}
Now, unconditioning with respect to $V$ in~\eqref{eq:ph2}, we obtain:
\begin{multline*}
  p_{in}(t\mid \tau)=\mathbb{E}_V\left[ 1-  e^{-V \int_ {t-T_C}^{t}\lambda(\theta-\tau)d\theta} \right]= \\
  1- {\phi_V \left(-\int_{t-T_C}^{t} \lambda(\theta-\tau)d\theta
    \right)}
\end{multline*}
To evaluate the probability of finding the given content in the cache
at time $t$, we uncondition with respect to $\tau$ and obtain:
\begin{equation}\label{eq:pin}
  p_{in}(t)=\frac{1}{t}\int_{0}^t
  1- {\phi_V \left(-\int_ {t-T_C}^{t} \lambda(\theta-\tau)d\theta \right)} d \tau  
\end{equation}
This result exploits the elementary property of Poisson processes
that, when a point falls within a specified interval of time, its
distribution is uniform over the considered interval.

Now, as in the case of the stationary popularity scenario, for a
sufficiently large $t$, the cache is completely filled with contents
introduced to the catalogue before $t$ and the number of contents in
the cache is exactly equal to the size of the cache. We can therefore
claim:
\[
C= \sum_m [\Uno_{\{\text{content m in cache}\mid \tau_m\le t\}}
\Uno_{\tau_m\le t}]
\] 
where $\tau_m$ is the time at which content $m$ is introduced into the
catalogue, and the sum extends over all the contents in the infinite
content catalogue.  Averaging both terms we obtain:
\begin{equation}\label{eq:c}
  C= \sum_m \mathbb{E} [ \Uno_{\{\text{content m in cache}\mid \tau_m\le t\}} \Uno_{\tau_m\le t}]= 
  p_{in}(t) \sum_m \mathbb{E}[\Uno_{\tau_m\le t}]
\end{equation}
Since the average number of contents introduced to the catalogue at
any time interval of size $t$ is $\gamma$, by combining~\eqref{eq:pin}
with~\eqref{eq:c}, we can evaluate the size of the cache $C$ as:
\begin{multline}\label{eq:cg}
\hspace{-5mm}
  C = \left(\sum_m \frac{\mathbb{E}[\Uno_{\tau_m<t}]}{t}\right)
  \int_{0}^t 1- \phi_V\left(-\int_ {t-T_C}^{t} \lambda(\theta-\tau)d\theta \right) d \tau  
  = \\ 
  \gamma \int_{0}^t 1- \phi_V \left(-\int_ {t-T_C}^{t} \lambda(\theta-\tau)d\theta \right) d \tau 
\end{multline}
Equation~\eqref{eq:cg} proves~\eqref{eq:c2}, and must be solved
(numerically) to evaluate the eviction time ($T_C$) given a cache size
of $C$.

Having defined $T_C$, we now return to evaluating the hit
probability for a given content with the parameters $(\tau_0,
V_0)$. By definition, the request for a given content at time $t$,
generates a hit at the cache iff the content is located on the
cache. Therefore, the probability of a hit  is given by: 
\[
 p_{hit}(t\mid \tau_0,V_0)= p_{in}(t\mid \tau_0,V_0)
\]
Now, to uncondition $ p_{hit}(t\mid \tau_0,V_0)$ with respect to $V_0$
and $\tau_0$, we have to consider that the probability with which
contents are requested by users is biased toward contents with higher
instantaneous popularity.  
Let $N(V_0,\Delta V,\tau_0,\Delta \tau_0)$ be the average number of contents
that have been generated during the interval $[\tau_0, \tau_0 +\Delta
  \tau_0)$ with request volume in $[V_0, V_0+ \Delta V_0)$. Now the
probability that a request arrives for one of the  contents  defined above  is 
\[
\dfrac{N(V_0,\Delta V,\tau_0,\Delta \tau_0)}{\gamma t}\times
\dfrac{ V_0 \lambda(t-\tau_0)}{\mathbb{E}[V]}
\]
where the second term represents the instantaneous rate originated by every considered content.
Thus,   recalling~\eqref{eq:ph2} we have:
\begin{multline*}
  p_{\text{hit}}(t)= \mathbb{E}_{\tau,V} \left[\frac{ V
      \lambda(t-\tau)}{\mathbb{E}[V]}
    p_{in}(t\mid \tau,V)\right]=\\
  \mathbb{E}_{V} \int_0^t \left[\frac{ V
      \lambda(t-\tau)}{\mathbb{E}[V]} \left( 1- e^{-V \int_
        {t-T_C}^{t}\lambda(\theta-\tau)d\theta}\right)\right]
  d\tau=\\
  \int_0^t\lambda(t-\tau) \mathbb E_V\left( \dfrac{V}{\mathbb
      E[V]}-\dfrac{Ve^{-V \int_
        {t-T_C}^{t}\lambda(\theta-\tau)d\theta}}{\mathbb E[V]}
  \right) d\tau=\\
  \int_{0}^t \lambda(t - \tau) \left(1-\frac{\phi_V' \left(-\int_
        {t-T_C}^{t} \lambda(\theta-\tau)d\theta
      \right)}{\mathbb{E}[V]} \right)d \tau
\end{multline*} 
By substituting $\alpha=t-\tau$ and $\beta=t-\theta$:

\begin{equation}\label{eq:ph1}
  p_{\text{hit}}(t)=
  \int_{0}^t \lambda(\alpha) \left(1-\frac{\phi_V' \
      \left(-\int_ {0}^{T_C} \lambda(\alpha-\beta)d\beta \
      \right)}{\mathbb{E}[V]} \right)d \alpha 
\end{equation} 
Thanks to the integrability property of $\lambda(t)$, (\ref{eq:phit})
is obtained by letting $t \to \infty$ in~\eqref{eq:ph1}.
\end{IEEEproof}

The following corollary sheds some light on the relation between $C$
and $T_C$:
\begin{corollario}\label{coroTC-C}
  The variables $C$ and $T_C$ satisfy:
  \begin{equation}\label{eq:ct}
    C  \le \gamma \mathbb E[V] \int_0^\infty \int_0^{T_C}
    \lambda(\tau-\theta) d\theta d\tau=\gamma \mathbb E[V] T_C
  \end{equation}
  and 
  \begin{equation}\label{eq:ct2}  
  C  \ge \gamma\mathbb E[V] T_C- \gamma\frac{\mathbb{E}[V^2]}{2} \int_0^\infty \left(\int_0^{T_C}
      \lambda(\tau-\theta) d\theta \right)^2 d\tau 
  \end{equation}
\end{corollario}
\IEEEproof The inequality in \eqref{eq:ct} is derived for
\eqref{eq:c2} by exploiting the inequality $1+x \le e^x$.  In
particular we exploit the previous inequality to lower bound
$\phi_V(-\int_0^{T_C}\lambda(\tau-\theta)d\theta)= \mathbb{E}
[e^{-\int_0^{T_C}\lambda(\tau-\theta)d\theta)}]$ with $1-
\mathbb{E}[V]\int_0^{T_C}\lambda(\tau-\theta) d\theta$ inside the
integral appearing in (\ref{eq:c2}).

Similarly, the inequality in \eqref{eq:ct2} is obtained by
exploiting $1-x+\frac{x^2}{2} \ge e^{-x}$ for every $x\geq 0$.
\endIEEEproof 

Note that the upper bound in~\eqref{eq:ct} has a simple meaning;
assuming that all the requests are referring to different contents,
the cache size is bounded by the overall number of requests $\gamma
E[V]$ during the time interval $T_C$. In Sec.~\ref{sec:sc}, we will
show that this bound is a good approximation for $C$ when cache size
is small.

To gather more insights on the impact of different parameters on
$p_{\text{hit}}$, we now derive a simplified expression for the two
extreme scenario regimes of {\em large} cache and {\em small} cache
sizes.

\subsection{Large-cache regime}\label{sec:hit:large}
A closed form expression for the asymptotic hit probability
($p_{\text{hit}, \infty}$) when the cache $C\to \infty$ 
can be derived from (\ref{eq:phit}) by making $T_C\to
\infty$.
\begin{corollario}
  For large cache sizes,
  \begin{equation}\label{eq:pinf}
    p_{\text{hit}, \infty}=1 - \frac{1-\phi_V(-1)}{\mathbb{E}[V]}=
    1-\dfrac{1}{\mathbb E[V]}+ \dfrac{\mathbb E[e^{-V}]}{\mathbb E[V]}
  \end{equation}
\end{corollario}
\begin{IEEEproof}
  Consider the limit as $T_C\to\infty$ in the integral
  within~\eqref{eq:phit}; it holds that:
  \begin{multline}\label{eq:1}
    \int_{0}^{\infty} \lambda(\tau)\phi_V'
    \left(-\int_0^\infty \lambda(\tau-\theta)d\theta\right)d\tau=\\
    \int_{0}^{\infty} \lambda(\tau)\phi_V'
    \left(-\int_0^\tau\lambda(\alpha)d\alpha\right)d\tau
  \end{multline}
  Now if we define $\Lambda(\tau)=\int_0^\tau\lambda(\alpha)d\alpha$
  (by construction, $\Lambda(\alpha)$ is also the primitive of
  $\lambda(\alpha)$) and $\beta=\Lambda(\tau)$, by substituting
  $\beta$ into~\eqref{eq:1}, we obtain:
  \begin{multline}\label{eq:2}
    \int_{0}^{\infty} \lambda(\tau)\phi_V'(-\Lambda(\tau))
    d\tau=
    \int_{0}^1\phi'_V(-\beta)d\beta=\\
    \phi_V(0)-\phi_V(-1)=1-\phi_V(-1)
  \end{multline} 
  Finally,~\eqref{eq:pinf} is obtained by using~\eqref{eq:2}
  within~\eqref{eq:phit}.
\end{IEEEproof}

Observe that~\eqref{eq:pinf} depends heavily on the distribution of
the content request volumes ($V$), and is completely independent of
the temporal profile of the popularity ($\lambda(t)$).  This is
expected, when we consider that as $C$ and $T_C$ grow large, contents
are never evicted from the cache. In effect, only the first request
for every content will lead to a cache miss, independently of the
arrival request pattern.

The expression (\ref{eq:pinf}) is exact, since for $C\to \infty$ it
can be easily proved that $T_C(m) \to \infty$ w.p.1.  This is obtained
exploiting by the following properties: i) as $C\to \infty$ the
conditional hit probability for contents originating $R\ge 1$ requests
tends to $p_{\text{hit}}(R)=1-{1}/{R}$, and ii) the probability of
observing at least one request for content $m$ is $\Pr(R\geq
1)=1-e^{V_m}$.


The value of $C$ (and consequently $T_C$) for which $p_{\text{hit}}$
approaches $p_{\text{hit}, \infty}$, instead heavily depends the
popularity profile $\lambda(t)$. Indeed it is possible to derive a
bound on the difference of the hit probability from
$p_{\text{hit}, \infty}$:

\begin{corollario}~\label{col:a1}
\vspace{-4mm}
  \[p_{\text{hit}, \infty}- p_{\text{hit},T_C} \le \int_{T_C}^{\infty}
  \lambda( \tau) d\tau \]
\end{corollario}
\begin{conference}
  The proof for Corollary~\ref{col:a1} is available in~\cite{techrep}.
\end{conference}
\begin{techrep}
  \begin{IEEEproof}
    Starting from~\eqref{eq:phit}, we obtain:
    \begin{multline}\label{eq:23}
      p_{\text{hit}}= 1-\int_0^{\infty}\lambda( \tau) \frac{\phi_V'
        \left(-\int_ {0}^{T_C} \lambda(\tau-\theta)d\theta
        \right)}{\mathbb{E}[V]}  d \tau  =\\
      \int_0^{\infty}\lambda( \tau)\left[1-\frac{\phi_V' \left(-\int_
            {0}^{T_C} \lambda(\tau-\theta)d\theta
          \right)}{\mathbb{E}[V]}\right]  d \tau  =\\
      \int_0^{T_C} \lambda( \tau) \left[1-\frac{\phi_V' \
          \left(-\int_ {0}^{\tau} \lambda(\alpha)d\alpha \right)}{\mathbb{E}[V]}\right]  d \tau +\\
      \int_{T_C}^{\infty} \lambda( \tau) \left[ 1-\frac{\phi_V'
          \left(-\int_ {0}^{T_C} \lambda(\tau-\theta)d\theta
          \right)}{\mathbb{E}[V]}\right] d \tau
\end{multline} 
where we have operated the change of variable $\alpha=\tau-\theta$.
By observing that 
\(
\phi'(x)=\mathbb E[Ve^{xV}]\leq \mathbb
E[V]
\) for any $ x\leq 0$, 
it is possible to upper bound the right-most term
of~\eqref{eq:23} as follows:
\begin{equation}\label{eq:5}
  \int_{T_C}^{\infty} \lambda( \tau)\left[1- \frac{\phi_V' \left(-\int_ {0}^{T_C} \
        \lambda(\tau-\theta)d\theta \right)}{\mathbb{E}[V]}\right] \
  d \tau \leq \int_{T_C}^{\infty} \lambda( \tau) d\tau
\end{equation}
Thanks to~\eqref{eq:2} it is also possible to upper bound the
left-most term:
\begin{multline}\label{eq:4}
  \int_0^{T_C} \lambda( \tau) \left[ 1-\frac{\phi_V' \left(-\int_
        {0}^{\tau} \lambda(\theta)d\theta \right)}{\mathbb{E}[V]}\right]
  d \tau \le \\
  \int_0^{\infty} \lambda( \tau) \left[1-\frac{\phi_V' \left(-\int_
        {0}^{\tau} \lambda(\theta)d\theta \right)}{\mathbb{E}[V]}\right]
  d \tau= \\ 
  1 - \int_0^\infty \lambda(\tau)\dfrac{\phi'_V(-\Lambda(\tau))}{\mathbb E[V]}d\tau
  =1-\dfrac{1-\phi_V(-1)}{\mathbb E[V]}
\end{multline}
which corresponds to $p_{\text{hit}, \infty}$. By combining~\eqref{eq:5}
with~\eqref{eq:4}, we get the assert.
 \end{IEEEproof}
\end{techrep}

\subsection{Small-cache regime}\label{sec:sc}
Under this regime, we get the following hit probability:
\begin{corollario}\label{coro:psmall}
  For very small cache sizes, we can approximate the hit
  probability as:
\begin{equation}\label{eq:psmall}
  p_{\text{hit}}\approx \frac{\mathbb E[V^2] }{\mathbb E[V]}T_C\int_0^\infty \lambda^2(\tau)d\tau
\end{equation}
\end{corollario}
\IEEEproof 
The expression in~\eqref{eq:psmall} is obtained from~(\ref{eq:phit})
by assuming $\int_0^{T_C} \lambda(\tau-\theta) d\theta \approx
\lambda(\tau)T_C \ll 1$, and approximating
$\phi_V'(x)=\mathbb{E}[Ve^{xV}] \approx \mathbb{E}[V] +x
\mathbb{E}[V^2]$ for small values of $x$.
\endIEEEproof

Furthermore, we can improve the results in Corollary~\ref{coroTC-C} to
better approximate the relation between $C$ and $T_C$ as follows:
\begin{equation}\label{eq:cts}
C\approx\gamma \mathbb E[V] T_C
\end{equation}

Following the same reasoning as the proof for~\eqref{eq:ct}, observe
that, for small values of $T_C$, $\int_0^{T_C}\lambda(\tau-\theta)
d\theta\ll 1$. Now $\phi_V(-\int_0^{T_C}\lambda(\tau-\theta) d\theta)$
can be approximated with $1-
\mathbb{E}[V]\int_0^{T_C}\lambda(\tau-\theta) d\theta$ and the desired
relation is obtained.

Using~\eqref{eq:cts}, we can now rewrite (\ref{eq:psmall}) as follows:
\begin{equation}\label{eq:psmall2}
  p_{\text{hit}}\approx \frac{\mathbb E[V^2]} { \mathbb {E}^2[V]}\dfrac{C}{\gamma}\int_0^\infty \lambda^2(\tau)d\tau
\end{equation}

\begin{techrep}
\begin{table}[tb]
  \centering
  \begin{tabular}{|l|c|c|}
    \hline  
    Profile & $\lambda(t)$ & \rule{0pt}{3mm} $\int_0^\infty \lambda^2(\tau)d\tau $\\
    \hline
    Exponential & $\frac{1}{L} e^{-t/L}$ for $t\geq 0$ & $\frac{1}{2L}$\\
    Power law ($\zeta>1$) &
    $\dfrac{\zeta-1}{L}\left(\dfrac{t}{L}+1\right)^{-\zeta}$ for $t\geq 0$& 
    $\dfrac{(\zeta-1)^2}{L(2\zeta-1)}$\\
    Uniform & $\frac{1}{2L}$ for $t\in[0,2L]$ & $\frac{1}{2L}$\\
    Triangular 
    & \rule{0pt}{6mm}$\begin{cases}
      \frac{t}{L^2} & \text{for}~t\in[0,L]\\
      \frac{2L-t}{L^2} & \text{for}~t\in[L,2L]\\
    \end{cases}$ 
    & $\frac{2}{3L}$ \\
    \hline
  \end{tabular}
  \caption{Examples  of popularity profiles $\lambda(t)$. 
    For all profiles the average content lifetime is set equal to $L$.
    Observe that  $\int_0^\infty \lambda^2(\tau)d\tau$ is the key parameter appearing in~\eqref{eq:psmall2}.}
  \label{tab:profiles}
\end{table}
\end{techrep}

\begin{conference}
\begin{table}[tb]
  \centering
  \caption{Examples  of popularity profiles $\lambda(t)$. 
   For all profiles the average content lifetime   is set equal to   $L$.
}\vspace{-3mm}
  \begin{tabular}{|l|c|c|}
    \hline  
    Profile & $\lambda(t)$ & \rule{0pt}{3mm} $\int_0^\infty \lambda^2(\tau)d\tau $\\
    \hline
    Exponential & $\frac{1}{L} e^{-t/L}$ for $t\geq 0$ & $\frac{1}{2L}$\\
    Power law ($\zeta>1$) &
    $\dfrac{\zeta-1}{L}\left(\dfrac{t}{L}+1\right)^{-\zeta}$ for $t\geq 0$& 
    $\dfrac{(\zeta-1)^2}{L(2\zeta-1)}$\\
    \hline
  \end{tabular}
\vspace{-5mm}
\label{tab:profiles}
\end{table}
\end{conference}

The expression given in~\eqref{eq:psmall2} enlightens us to the
potentially large effect the popularity profile of content has on the
effectiveness of caching, when cache sizes are small. For
illustration, if we consider the popularity profiles given in
Table~\ref{tab:profiles}, the hit probability is always inversely
proportional to the average content lifetime ($L$). But, by
comparing the third column, it is clear that the actual value depends
strongly on the shape of the profile.


\section{Numerical validation}\label{sec:validation}
In this section we present: i) the results of applying the
approximation of LRU under non-stationary traffic (see
Sec.\ref{sec:traffic}), as given by Theorem~\ref{theo:1}; ii) the
validation of the predictions of the model through Monte Carlo
simulations of a single cache.



The results presented in this section relate the size of a cache ($C$)
to the hit probability ($p_{\text{hit}}$) and look at relation
between these two variables, when varying: i) the average content
lifetime ($L$), ii) the average content request volume ($\mathbb{E}[V]$), iii)
the  distribution of content request volumes,  
iv) the shape of the popularity
 time profile. 

 For all the results that follow, we set the content arrival rate
 ($\gamma$) to $10$k contents per day and the content request volume
 ($V$) is assumed to be distributed according to a Pareto
 distribution: $f_V(v)=\beta V_{\min}^\beta/v^{1+\beta}$ for
   $v\ge V_{\min}$. The choice of a Pareto distribution is justified
 by two factors. First, several recent measurement studies have shown
 that the Zipf law is a very good approximation of the empirical
 distribution of long term content (video) request
 volumes~\cite{Roberts12,zipf}.  Second, a Zipf-like distribution of
 content request volumes with parameter $\alpha={1}/({\beta-1})$ is
 obtained when a large number of individual content request volumes
 are independently generated according to Pareto distribution.

\begin{techrep}
  With regard to the different popularity profiles that content may
  display, we consider exponential and power law profiles as given in
  Table~\ref{tab:profiles}.
\end{techrep}
\begin{conference}
  With regard to the different popularity profiles that content may
  display, we consider exponential and power law profiles as given in
  Table~\ref{tab:profiles}.
\end{conference}
Finally, the parameter $\zeta$ is used to model the different 
time-dependent popularity profiles (popularity shapes).

Figs.~\ref{fig:phit_vs_cachesize_exp_mod-sim_log}
and~\ref{fig:phit_vs_cachesize_powlaw_mod-sim_log} report the hit
probability for different values of content lifetime, with respect to using; i)
an exponential (see Fig.~\ref{fig:phit_vs_cachesize_exp_mod-sim_log})
and ii) power law with parameter $\zeta=3$ (see
Fig.~\ref{fig:phit_vs_cachesize_powlaw_mod-sim_log}) popularity
profile.  In both cases, we set $V_{\min}=1$ and $\beta=3$, as a
consequence, we obtain $\mathbb{E}[V]=1.5$ requests per content.

The first point to observe is that the model estimates agree strongly
with the simulation results for the hit probability in all the
cases. Second, as predicted by the model, the average content lifetime
($L$) deeply impacts the cache performance.  Indeed, for a given cache
size ($C$), a given content's hit probability increases significantly
as its lifetime is reduced.  In particular, for moderate cache sizes,
the hit probability is roughly inverse proportional to the content
lifetime, as predicted by Corollary~\ref{coro:psmall}.

\tgifeps{8}{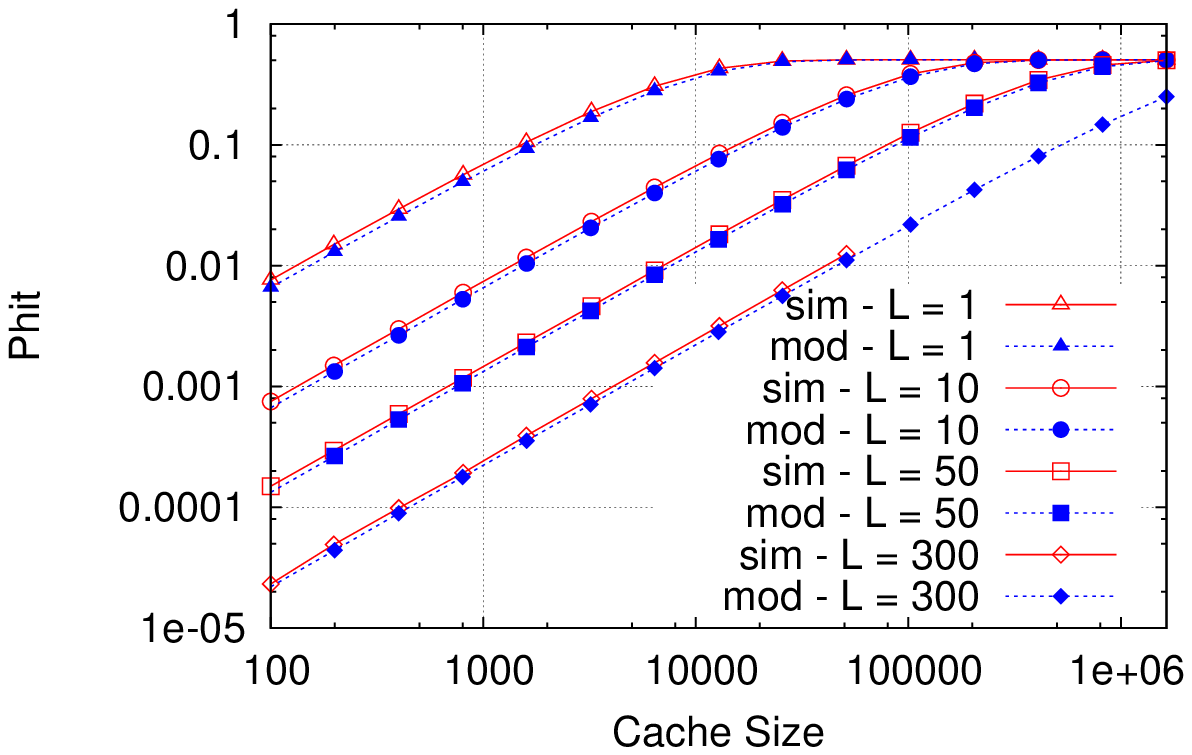}{Cache hit probability
  under exponential popularity profile, for different values of the
  average content lifetime $L$ (expressed in days).}

\tgifeps{8}{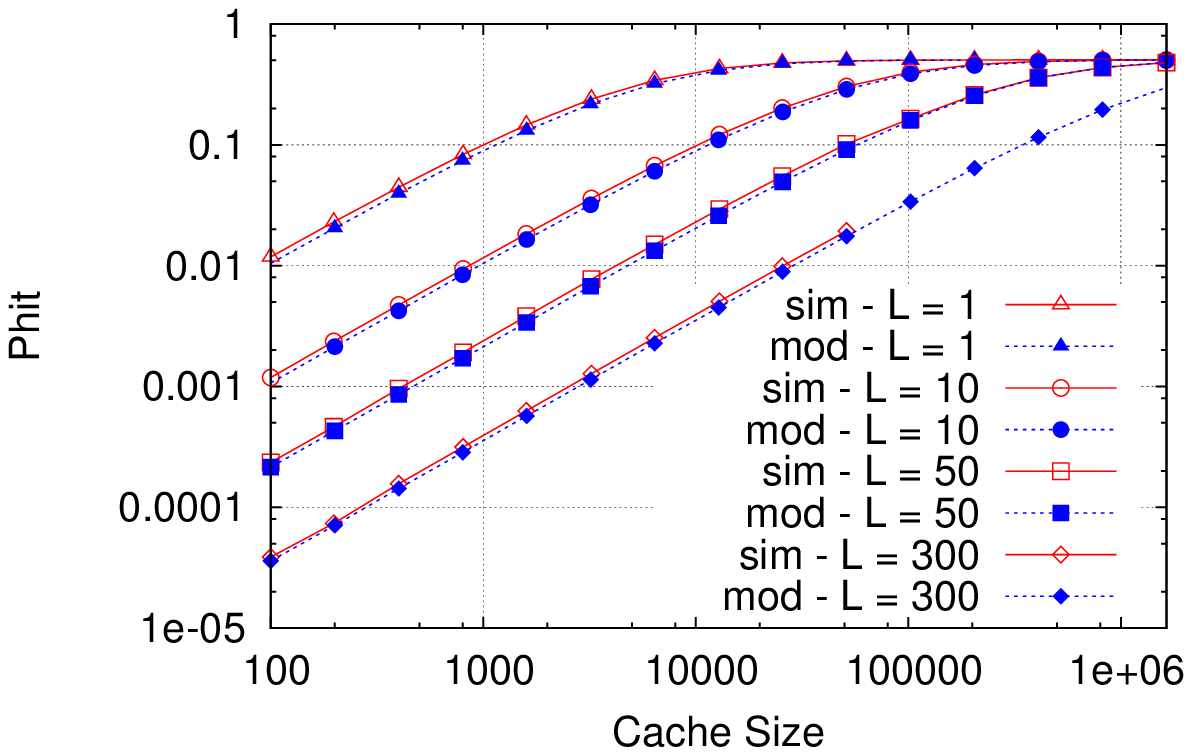}{Cache hit probability
  under power law popularity profile with $\zeta=3$, for different
  values of the average content lifetime $L$ (expressed in
  days).}

Fig.~\ref{fig:phit_vs_cachesize_exp_meanpop_mod-sim} reports the cache
hit probability for different values of the average content request volume
$\mathbb{E}[V]$. All plots refer to the same value of $\beta=3$ and
different values of $V_{\min}=\mathbb{E}[V]\frac{\beta-1}{\beta}$. The
results here reveal that the volume of hits accumulated by content has
impact exclusively for large cache sizes, i.e. when $C \ge 200$k
objects, and as predicted by~\eqref{eq:pinf}, the cache hit
probability increases with the volume of content requests. For
moderate cache sizes ($C \le 10$k), the effect of the hits
becomes negligible as predicted by (\ref{eq:psmall2}). Indeed, all
curves correspond to the same value of
$\frac{\mathbb{E}[V^2]}{\mathbb{E}^2 [V]}$.

\tgifeps{8}{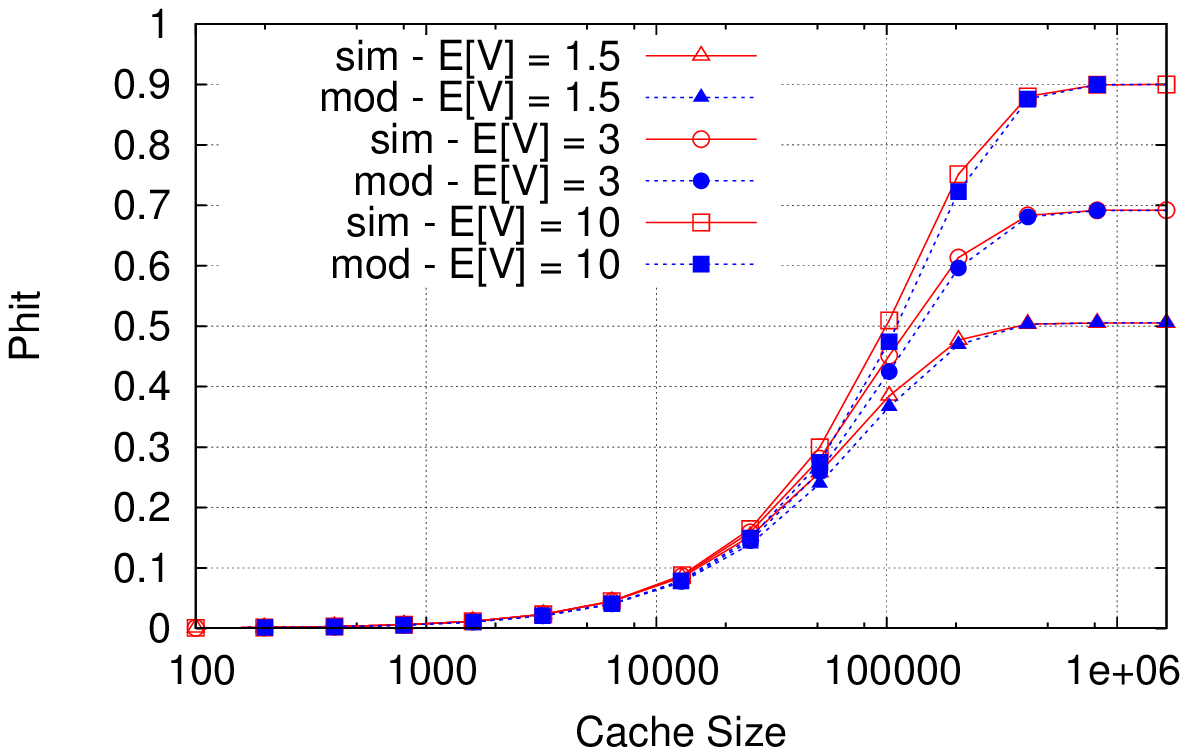}{Cache hit probability
  under exponential popularity profile with $L=10$ days for different
  values of the average content volume $E[V]$.}

Fig.~\ref{fig:phit_vs_cachesize_exp_beta_mod-sim_log2} reports the
cache hit probability for different values of the parameter $\beta$,
associated to the distribution of content request volumes.  From the
figure we see that the shape of requests volumes has a significant
impact on the cache hit probability.  As expected, by decreasing
$\beta$ (i,e., increasing the correspondent parameter $\alpha$ for the
associated Zipf law), the cache performance is improved. In particular
we observe that the caching performance become much more sensitive to
$\beta$ as $\beta$ gets smaller than $2$ (i.e., as the corresponding
Zipf parameter $\alpha$ increases above $1$).  However, the impact of
$\beta$ (i.e., $\alpha$) on caching performance does not appear in our
scenario to be as strong as it does in the classical stationary case,
where a sort of ``phase transition'' is observed as $\alpha$ crosses
the value $1$~\cite{Roberts12}.


\tgifeps{8}{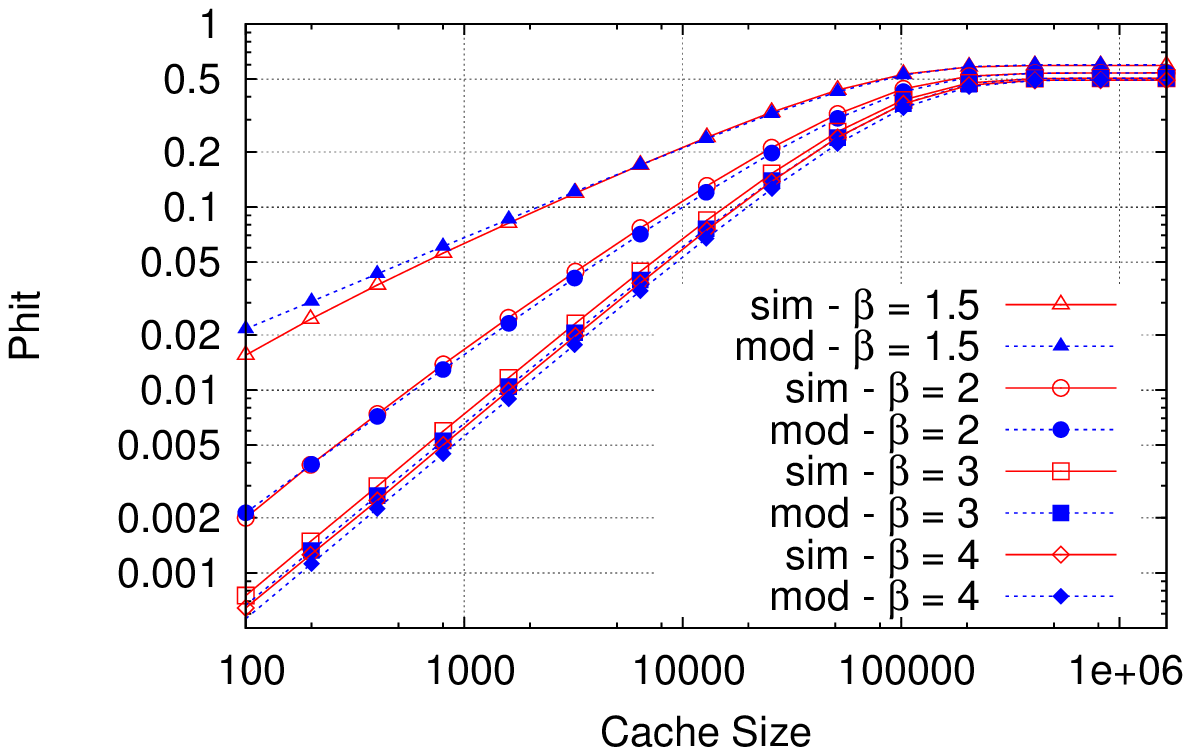}{Cache hit
  probability under exponential popularity profile for different request
  volume distributions and average content request volume $E[V]=1.5$.}

Finally, Fig.~\ref{fig:phit_vs_cachesize_powlaw_zeta_mod-sim_log2}
reports the cache hit probability for different content popularity
profiles (i.e, varying $\zeta$) - while keeping the average lifetime as
constant at $L=10$ days.  From the figure, we see that the content
popularity profile appears to have a moderate impact on the caching
performance (for small caches) as long as the average content
lifetime $L$ is kept constant. For the extreme case where the size
cache $C=100$, $p_{\text{hit}}$ varies from $0.0032$ and $0.001$ as
$\zeta$ is decreased from $4$ to $2.2$ (see the simulation curves).

\tgifeps{8}{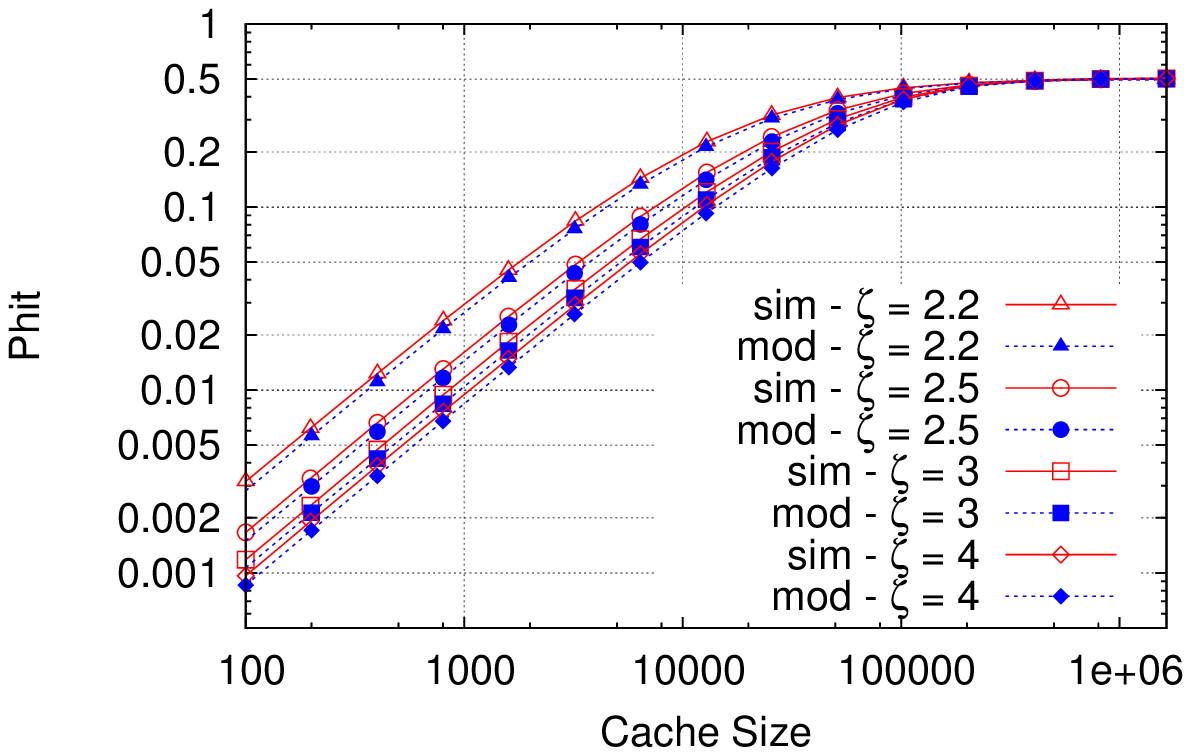}{Cache hit
  probability under power law popularity profile for different values
  of $\zeta$ and the same average content lifetime $L=10$ days.}

\section{Extension to a Multi-class scenario}~\label{sec:ext}
In this section, we consider a more realistic scenario in which
contents can be partitioned into $K$ classes, such that each class 
is associated with a different popularity profile
$\lambda_k(t)$ and a different request volume distribution $V_k$,
for $1\le k\le K$.

This generalisation of our traffic model is needed in order to capture
the variability of the popularity profiles exhibited by real contents.
Indeed the popularity profile depends heavily on the nature of
contents, for example, the popularity evolution of music videos is
typically significantly different to videos containing sport
highlights. However, recent experimental studies~\cite{Ahmed:2013}
have shown that the popularity evolution of different contents can be
clustered to relatively few groups exhibiting similar temporal
popularity profiles.

We can formalise the multi-class scenario by assuming that every
generated content ($m$) is associated with a random mark $W_m$ taking
values in $\{1, \ldots, K\}$, such that the mark specifies the class
the content belongs to.  Assuming $\{W_m \}$ to be i.i.d.\ random
variables, the total stochastic intensity at time $t$ of the request
process is given by:
\[
\Lambda(t)=\sum_{m} V_m \lambda_{W_m}(t-\tau_m)=\sum_{m} V_m
\lambda_{W_m}(t-\tau_m)\Uno_{\{\tau_m\le t \}}
\]
Under this assumption, we can now state the following:
\begin{teorema}~\label{theo:mcLRU}
  Consider a cache of size $C$ implementing LRU policy, 
  operating under a multi-class non-stationary popularity model, with
  total stochastic intensity: \( \Lambda(t)=\sum_{m} V_m
  \lambda_{W_m}(t-\tau_m) \). Extending Che's approximation, 
  the hit probability is given by:
\[
\hspace{-5mm}
p_{\text{hit}}=1- \sum_{k=1}^K\Prob\{W_1=k\} \int_0^{\infty}
\lambda_k( \tau) \frac{\phi_{V_k}'\left(-\int_ {0}^{T_c}
    \lambda_k(\tau-\theta)d\theta \right)}{\mathbb{E}[V_k]}d \tau
\]
where $T_C$ is the solution to the equation:
\[
\hspace{-5mm}
C= \gamma \int_{0}^{\infty}\left[ 1- \sum_1^K
  \Prob\{W_1=k\}{\phi_{V_k}\left(-\int_ {0}^{T_C}
      \lambda_k(\tau-\theta) d\theta \right)}\right] d \tau
\]
\end{teorema}
The proof for Theorem~\ref{theo:mcLRU} (not reported here for the sake
of brevity) follows the same lines of Theorem~\ref{theo:1}.

\tgifeps{8}{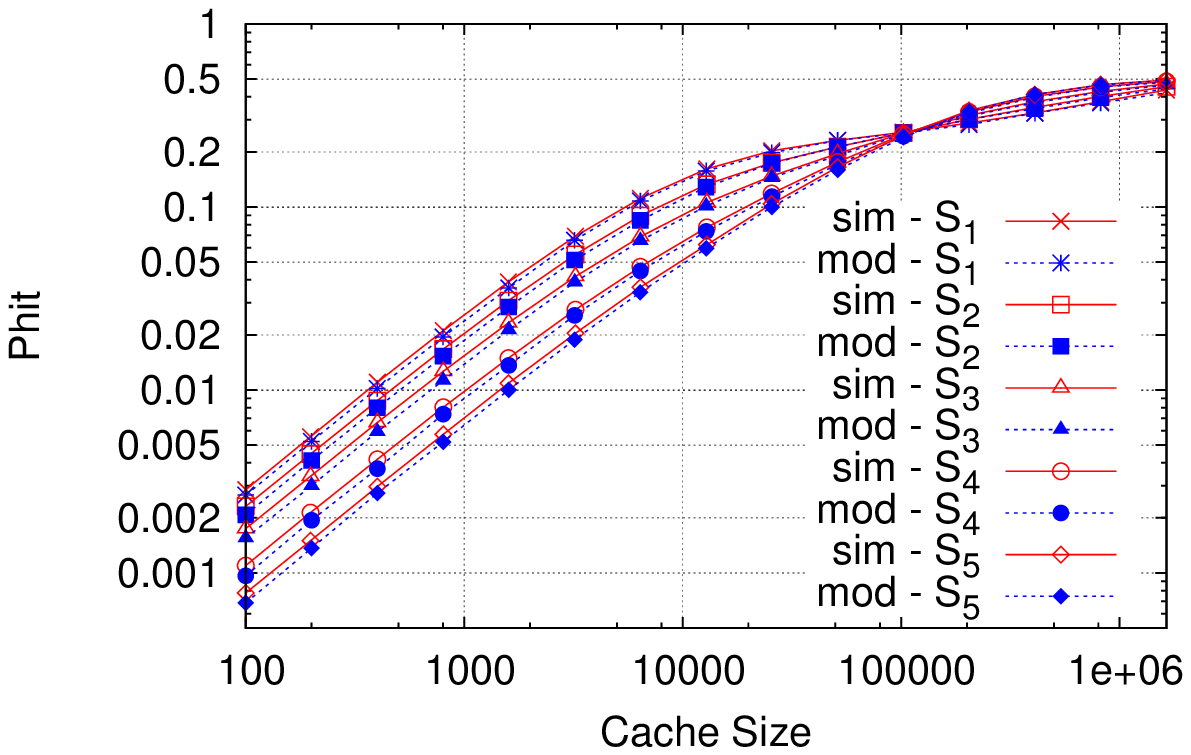}{Cache hit
  probability under exponential popularity profile for different
  classes configurations.}

\begin{table}[tb]
\centering
\small
\begin{tabular}{|c|c|c|c|c|c|c|}\hline
Popularity class    	&$L$ [days]  	& $S_1$ & $S_2$ & $S_3$ & $S_4$ & $S_5$	\cr\hline
1      			&1        	& 40\%	& 30\%	& 20\%  & 10\% 	& 5\%  	\cr
2      			&10        	& 10\%	& 20\%	& 30\%  & 40\% 	& 45\%	\cr
3      			&50        	& 10\%	& 20\%	& 30\%  & 40\% 	& 45\%	\cr
4      			&300       	& 40\%	& 30\%	& 20\%  & 10\% 	& 5\%  	\cr\hline
\end{tabular}
\caption{Average content lifetime $L$ and fractions of
  contents in each class for setups from $S_1$ to $S_5$.}
\label{tab:multiclass} 
\vspace{-11mm}
\end{table}
In order to evaluate the multi-class scenario, we partition contents
into $K=4$ classes, each characterised by a different popularity
profile and lifetime. For simplicity, here, we only considered 5
different configurations as specified in Table~\ref{tab:multiclass}.

Fig.~\ref{fig:phit_vs_cachesize_exp_multiclass_mod-sim_log2} reports
the cache hit probability for each setup.  First, from the figure, we
see that the model predictions of the cache hit probability for
different cache sizes align accurately with the results of the
simulation.  Second,
Fig.~\ref{fig:phit_vs_cachesize_exp_multiclass_mod-sim_log2} shows
that the heterogeneity of contents weakly impacts the cache
performance, and that the cache hit probability increases when the
average content lifetime ($L$) of each setup decreases, i.e.  when the number
of contents with fast popularity dynamics (belonging to class 1) is
large with respect to the rest. In fact, as already seen in
Fig.~\ref{fig:phit_vs_cachesize_exp_mod-sim_log}, contents attracting
users' attention for very limited periods of time ($L=1$ days) are
largely responsible for improving the cache performance when adopting
the LRU strategy.
\vspace{-2mm}
\section{Conclusions}
This work has proposed (and validated with a Monte Carlo simulation) a
simple but highly accurate approximated model for an LRU (Least
Recently Used) cache under non-stationary traffic conditions. The
proposed model is flexible and can be easily extended to consider
complex and realistic traffic scenarios. It has the advantage of being
computationally cheap when compared with the Monte Carlo simulations -
especially when cache sizes are large.

Our results show that caching performance is deeply impacted by the
dynamics resulting from the popularity of content. In particular, we
find that when cache sizes are small as in the case anticipated for
ICN networks, the content hit probability is largely sensitive to the
popularity profile of contents.





\begin{techrep}
\appendices
\section{LRU under stationary popularity}\label{sec:static}

First we briefly resume Che's approximation of LRU in a classical
traffic scenario in which every content $m$, in a finite catalog of
size $M$, presents a time-invariant popularity profile.  More in
details, we assume that requests for content $m$ arrive at the cache
according to a time homogeneous Poisson process with intensity
$\lambda_m$.  Let $\Lambda= \sum_m \lambda_m $ be the resulting total
arrival intensity of content requests at the cache.

Now, thanks to Che's approximation, discussed in Sec.~\ref{sec:hit}, a
content $m$ is present in the cache at time $t$, if and only if a time
less than $T_c$ has passed since the last request for content $m$,
i.e., if at least a request for such content has arrived in the
interval $(t-T_c,t]$.  Since requests arrivals are Poisson, the
probability $p_h(m)$ that at at least one request has arrived in the
interval $(t-T_c,t]$ is given by: \( 
p_h(m) = 1- e^{-\lambda_m T_c} \). 
Observe that $p_h(m)$ represents, by construction, also, the hit
probability for content $m$, as immediate consequence of PASTA
property of arrivals.

Considering a cache of size $C$, by construction: \( C= \sum_m
\Uno_{\{\text{$m$ in cache}\}} \).  When averaging both sides, we
obtain:
\[
C= \sum_m \mathbb{E}[\Uno_{\{\text{$m$ in cache}\}} ]= \sum_m p_h(m)=
\sum_m (1- e^{-\lambda_m T_c}).
\] 
By solving the above relationship, we obtain $T_c$, and then the
average hit probability on the cache as:
\begin{equation}\label{eq:phm}
p_{\text{hit}}= \sum_m p_m p_h(m)
\end{equation}

\subsection{Numerical evaluation}
We test the Che's approximation in the stationary popularity case.  We
set a catalog size equal to $M=10^7$ contents.
Fig.~\ref{fig:phit_vs_cachesize_stationary_log} shows the impact of
the Zipf's exponent $\alpha$ on the caching performance.

\tgifeps{9}{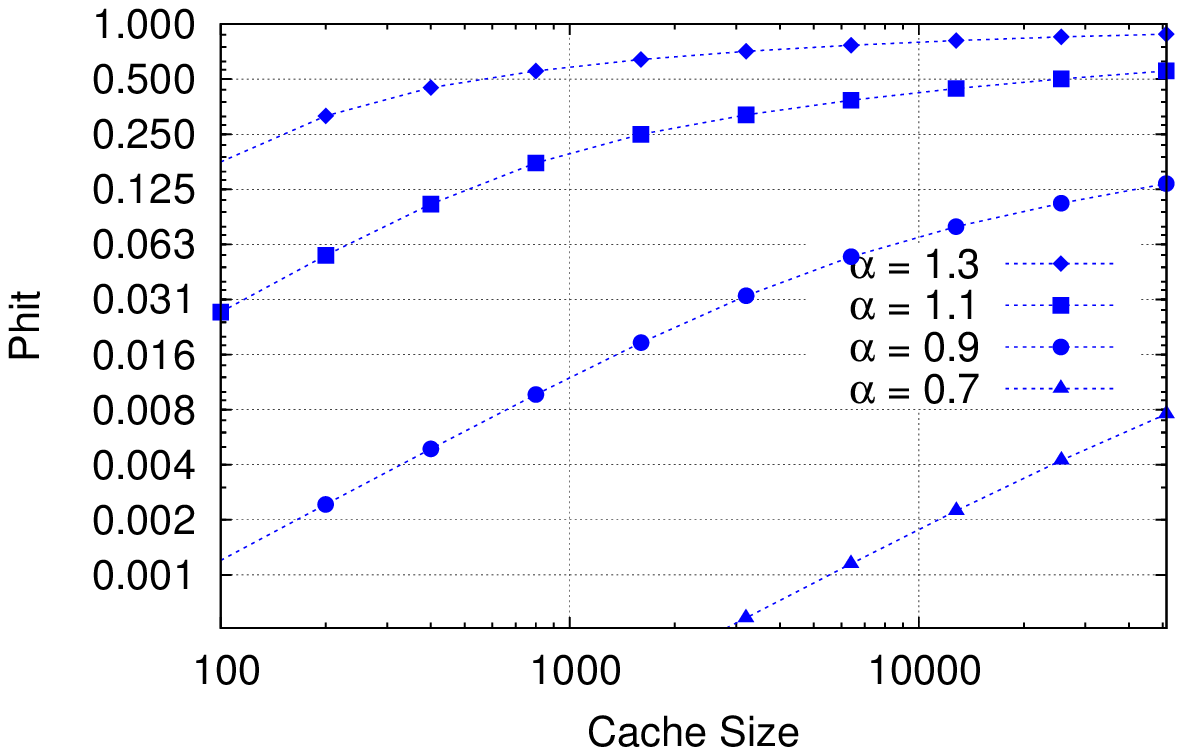}{Hit probability
  $p_{\text{hit}}$ vs.\ cache size for different content popularity
  distributions}

As already well known, the popularity distribution shape has a
disruptive impact on caching performance.

For $\alpha$ sufficiently larger than 1, the popularity distribution
is sufficiently skewed, so that the contribution of the few top
popular contents to the total traffic is significant.  Indeed, observe
that that $H(M)=\Theta(1)$ when $M$ grows large, i.e.  the aggregate
contribution of tail contents is marginal.

For $\alpha<1$, instead, only caching a significant portion of the
huge catalog we can achieve significant cache hit probability. In this
case $H(M)$ diverges as $M \to \infty$ showing that the aggregate
contribution of tail contents is dominant.

\end{techrep}
\end{sloppypar}

\end{document}